\documentclass[fleqn,usenatbib]{mnras}

\usepackage{newtxtext,newtxmath}

\usepackage[T1]{fontenc}

\DeclareRobustCommand{\VAN}[3]{#2}
\let\VANthebibliography\thebibliography
\def\thebibliography{\DeclareRobustCommand{\VAN}[3]{##3}\VANthebibliography}

\usepackage{graphicx}   
\usepackage{amsmath}    
\usepackage{CJKutf8}    
\usepackage{upgreek}    
\graphicspath{{figures/},{pics/}}
\usepackage{hyperref}   

\title[GWs from close-in strange quark planetary systems]{Gravitational
Wave Emission from Close-in Strange Quark Planets Around Strange
Stars with Magnetic Interactions}

\author[X.-L. Zhang et al.]{
Xiao-Li Zhang,$^{1}$
Ze-Cheng Zou,$^{2}$
Yong-Feng Huang,$^{2,3}$\thanks{E-mail: hyf@nju.edu.cn (Y-FH)}
Hao-Xuan Gao, $^{4}$
Pei Wang, $^{5,6}$
Lang Cui, $^{7}$
\newauthor and Xiang Liu$^{7}$
\\
$^{1}$Department of Physics, Nanjing University, Nanjing 210093, China\\
$^{2}$School of Astronomy and Space Science, Nanjing University, Nanjing 210023, China\\
$^{3}$Key Laboratory of Modern Astronomy and Astrophysics (Nanjing University), Ministry of Education, Nanjing 210023, China\\
$^{4}$Purple Mountain Observatory, Chinese Academy of Sciences, Nanjing 210023, China\\
$^{5}$National Astronomical Observatories, Chinese Academy of Sciences, Beijing 100101, China\\
$^{6}$Institute for Frontiers in Astronomy and Astrophysics, Beijing Normal University, Beijing 102206, China\\
$^{7}$Xinjiang Astronomical Observatory, Chinese Academy of Sciences, 150 Science 1-Street, Urumqi 830011, China
}

\date{Accepted XXX. Received YYY; in original form ZZZ}

\pubyear{2024}

\begin{document}
\label{firstpage}
\pagerange{\pageref{firstpage}--\pageref{lastpage}}
\maketitle

\begin{abstract}
According to the strange quark matter hypothesis, strange planets may exist, which are planetary mass objects composed of almost equal numbers of up, down and strange quarks. A strange planet can revolve around its host strange star in a very close-in orbit. When it finally merges with the host, strong gravitational wave emissions will be generated. Here the gravitational waveforms are derived for the merging process, taking into account the effects of the strange star's magnetic field on the dynamics. Effects of the inclination angle are also considered. Templates of the gravitational waveforms are derived. It is found that the magnetic interactions significantly speed up the merging process. Coalescence events of such strange planetary systems occurring in our Galaxy as well as in local galaxies can be effectively detected by current and future gravitational experiments, which may hopefully provide a new method to test the strange quark matter hypothesis and probe the magnetic field of compact stars.
\end{abstract}




\section{Introduction}

The detection of the binary black hole merger event GW150914 by the LIGO--Virgo Collaboration \citep{Abbott_GW150914} marks the beginning of gravitational-wave (GW) astronomy. The following years witnessed an incredibly increasing number of detected GW events from binary black hole mergers \citep{Abbott_GW151226,Abbott_GW170104,Abbott_GW170814,Abbott_GW170608,Abbott_GW190521}, binary neutron star mergers \citep{Abbott_GW170817,Abbott_GW190425}, and even neutron star--black hole mergers \citep{Abbott_GW190814,Abbott_GW200105_GW200115}. A larger range of astrophysics has become accessible through GWs, which provide new insights into the nature of the Universe. Especially, GWs are found to be able to help probe the interiors of compact stars thanks to the detection of GW170817 \citep[for a review, see][]{2019JPhG...46l3002G}.

Nevertheless, the exact composition and structure of compact stars are still elusive after huge efforts \citep[for a recent review, see][]{2021Univ....7..267M}. According to the strange quark matter (SQM) hypothesis \citep{Farhi&Jaffe_1984,Witten_1984}, nuclei consisting of baryons will undergo a phase transition and become a mixture composed of three quark flavors (up, down, and strange quarks) when they are subject to an extremely high pressure. Consequently, strange quark stars can stably exist and pulsars may actually be strange stars \citep{Itoh_1970,Alcock_1986_ApJ,1986A&A...160..121H,Geng_2021_Innov}. GW emissions from merging binary strange stars should be different from that of merging binary neutron stars \citep{Limousin_2005_PhRvD,Zhu&Rezzolla_2021_PhRvD}, but the difference is somewhat subtle and difficult to be discerned by current GW experiments \citep{2019JPhG...46k4001A}.

The self-bound nature of SQM means that strange quark nuggets would be stable. Consequently, SQM objects of planetary mass, i.e. strange planets, can also stably exist. Therefore, searching for strange planets would provide a direct test for the SQM hypothesis \citep{Geng_2015_Apjl,2017ApJ...848..115H,Kuerban_2019,Kuerban_2020}. Various processes can lead to the formation of strange planets around compact stars \citep{2006APh....25..212X,2012RAA....12..813H}. A distinct feature of strange planets is that they can be very close to their hosts. A normal planet will be tidally disrupted if its orbit period ($T$) is less than $\sim 6100\,\mathrm{s}$, while a strange planet can still safely exist even for a much smaller period due to its high density. Such a period criterion can be used to distinguish strange planets from normal matter ones \citep{Geng_2015_Apjl,2017ApJ...848..115H,Kuerban_2019,Kuerban_2020,2023mgm..conf.3118W}, although searching for such small period pulsar planets through timing observations is still challenging \citep{2018exha.book.....P}.

A close-in strange planet--compact star system can emit very
strong GWs, while a normal planet will be tidally disrupted far
before the GW emission becomes significant.
\citet{Geng_2015_Apjl} studied the GW emissions from
close-in SQM planetary systems and argued that such GW signals can
be used to probe SQM objects. It is found that a heavier strange
quark planet will lead to a quicker merging process associated
with a higher GW amplitude \citep{Geng_2015_Apjl}.
\citet{Kuerban_2020} further investigated the GW emissions from
ten candidates of close-in SQM planet systems in which the masses
of the strange quark planets are different. They argued that the
merger-induced GW emissions can be potentially detected by the
advanced LIGO and Einstein Telescope (see Figure 4 of
\citet{Kuerban_2020}). It is worth noting that in these previous
studies \citep{Geng_2015_Apjl,Kuerban_2019,Kuerban_2020},
researchers have mainly concentrated on the impact of planet mass
on GW emissions. However, a compact star usually has a strong
magnetic field, which will interact with the strange planet and
change the dynamics of the system. Therefore, in this
study, we will consider the effects of magnetic field and solve
the problem numerically to obtain a set of GW templates for
merging strange planet-compact star systems.

The structure of our paper is organized as follows. The magnetic
field interactions and the GW emission process are modeled in
Section~\ref{sec:method}. Our numerical results and the GW
templates are presented in Section~\ref{sec:result}. Finally,
Section~\ref{sec:ending} presents our conclusions and some brief
discussion.


\section{Model}
\label{sec:method}

\subsection{Magnetic Interactions}

For a rotating strange star, the radius of the light cylinder is mainly determined by its rotation speed, i.e.
\begin{equation}
    r_c=\frac{c}{\Omega}=6.7\times10^6\left(\frac{\nu}{716\,\mathrm{Hz}}\right)^{-1}\mathrm{cm},
    \label{eq:rc}
\end{equation}
where $c$ is the speed of light, $\Omega$ and $\nu$ are the angular rotation velocity and rotation frequency of the strange star, respectively \citep{Lyne_Graham-Smith_Stappers_2022}. The faster a compact star rotates, the smaller its light cylinder radius will be. Till now, the fastest rotating millisecond pulsar ever observed has a rotation frequency of $\nu=716\,\mathrm{Hz}$ \citep{2006Sci...311.1901H}. Its light cylinder radius is as small as $6.7 \times 10^6$ cm, but is still much larger than the typical radius of neutron stars and strange stars.

The strange planet will directly interact with the strange star's magnetosphere when its orbit is inside the light cylinder. Note that the tidal disruption radius of the strange planet is \citep{Geng_2015_Apjl,2017ApJ...848..115H}
\begin{equation}
    r_\mathrm{td} = 1.5 \times 10^6 \left( \frac{M}{1.4\,M_\odot} \right)^{1/3} \left( \frac{\rho}{4\times10^{14}\,\mathrm{g\,cm^{-3}}} \right)^{-1/3}\mathrm{cm},
    \label{eq:tidal}
\end{equation}
where $M$ is the mass of the strange star and $\rho \sim 4 \times 10^{14} \,\mathrm{g\,cm^{-3}}$ is the mean density of the strange planet. We see that $r_\mathrm{td}$ is only slightly larger than the radius of the strange star, which means the strange planet can maintain its integrity even when it is very close to the strange star surface. Combining Equations~(\ref{eq:rc}) and (\ref{eq:tidal}), we can safely regard the strange planet as an intact sphere when it interacts with the magnetosphere of the strange star inside the light cylinder.

When orbiting around the strange star inside the magnetosphere, the strange planet will travel across the magnetic field lines which gives birth to a strong electric field. In the magnetosphere that is full of free electrons, this will lead to strong electric currents and correspondingly, a strong electromotive force exerted on the strange planet itself. According to the popular unipolar induction direct current (DC) model \citep{Piddington_1968_Nature,Goldreich&Lynden-Bell_1969_ApJ}, the kinetic energy of the planetary system will be dissipated through Ohmic dissipation, which is determined by the total resistance of the circuit. In our framework, the surface of a bare strange planet is covered by a layer of free electrons \citep{Alcock_1986_ApJ}. Therefore, the strange planet can be regarded as a superconductor. \citet{Lai_2012_ApJ} proved that the Ohmic dissipation will not grow up to infinity even when the total resistance is extremely small. Instead, it has an upper limit because a large current will twist the magnetic flux tubes and destroy the circuit. As a result, the energy loss rate of the system due to the magnetic interaction is \citep{Lai_2012_ApJ}
\begin{equation}
    \dot{E}_\mathrm{mag} = -\zeta_\phi(\omega-\Omega) \frac{\mu^2 r^2}{2a^5},
    \label{eq:Emag}
\end{equation}
where $0 < \zeta_\phi < 1$ is a coefficient describing the twist of the magnetic flux tube, $\mu$ is the magnetic moment of the strange star, $\omega$, $r$, and $a$ are the strange planet's orbital angular velocity, radius, and orbit radius, respectively. According to the refined direct circuit model proposed by \citet{Lai_2012_ApJ}, the maximum energy dissipation, i.e. the smallest total resistance limit, occurs when $\zeta_\phi=1$.

\subsection{Dynamics and GW Emissions}

In addition to the energy dissipation from magnetic interactions, the system is also subjected to energy dissipation caused by GW emissions. To analyze the GW effects on the dynamics, we assume a quasi-circular orbit for the system for simplicity. Denoting the strange planet's mass as $m$, energy loss rate due to GW emissions can be expressed as \citep{Landau_1975_ctf..book,Jolien_2011_book,Postnov&Yungelson_2014_LRR}
\begin{equation}
    \dot{E}_\mathrm{GW} = - \frac{32G^4M^2m^2(M+m)}{5c^5a^5},\label{eq:Egw}
\end{equation}
where $G$ is the gravitational constant.

The dynamical evolution of the system can be solved numerically by combining Equations~(\ref{eq:Emag}) and~(\ref{eq:Egw}). After a small time interval $\updelta t$, the loss of the orbital energy is
\begin{equation}
    \updelta\left(-\frac{GMm}{2a}\right)=\left(\dot{E}_\mathrm{mag}+\dot{E}_\mathrm{GW}\right)\updelta t.
\end{equation}
The orbit will shrink as time goes on. The orbital phase at time $t$ can then be calculated as
\begin{equation}
    \varphi = \int_0^t\omega(t)\,\mathrm{d}t = \int_0^t \frac{2\pi}{T(t)} \, \mathrm{d}t,
\end{equation}
where $T$ is the orbital period.

After solving the dynamics, we can easily calculate the waveform of the emitted GWs. For an observer at a viewing angle of $\iota$ (the inclination angle with respect to the normal of the orbital plane), the waveform can be expressed as \citep{Jolien_2011_book}
\begin{equation}\begin{split}
        h_+&=-\frac{4Gua^2\omega^2}{c^4d}\frac{1+\cos^2\iota}{2}\cos2\varphi,\\
        h_\times&=-\frac{4Gua^2\omega^2}{c^4d}\cos\iota\sin2\varphi,\label{eq:GW}
\end{split}\end{equation}
where $h_+$ and $h_\times$ denote the two polarizations of GWs, $u = M m / (M+m)$ is the reduced mass of the strange planet-strange star system, and $d$ is the luminosity distance.

\begin{figure}
    \centering
    \begin{minipage}{0.49\textwidth}
        \centering
        \includegraphics[width=\columnwidth]{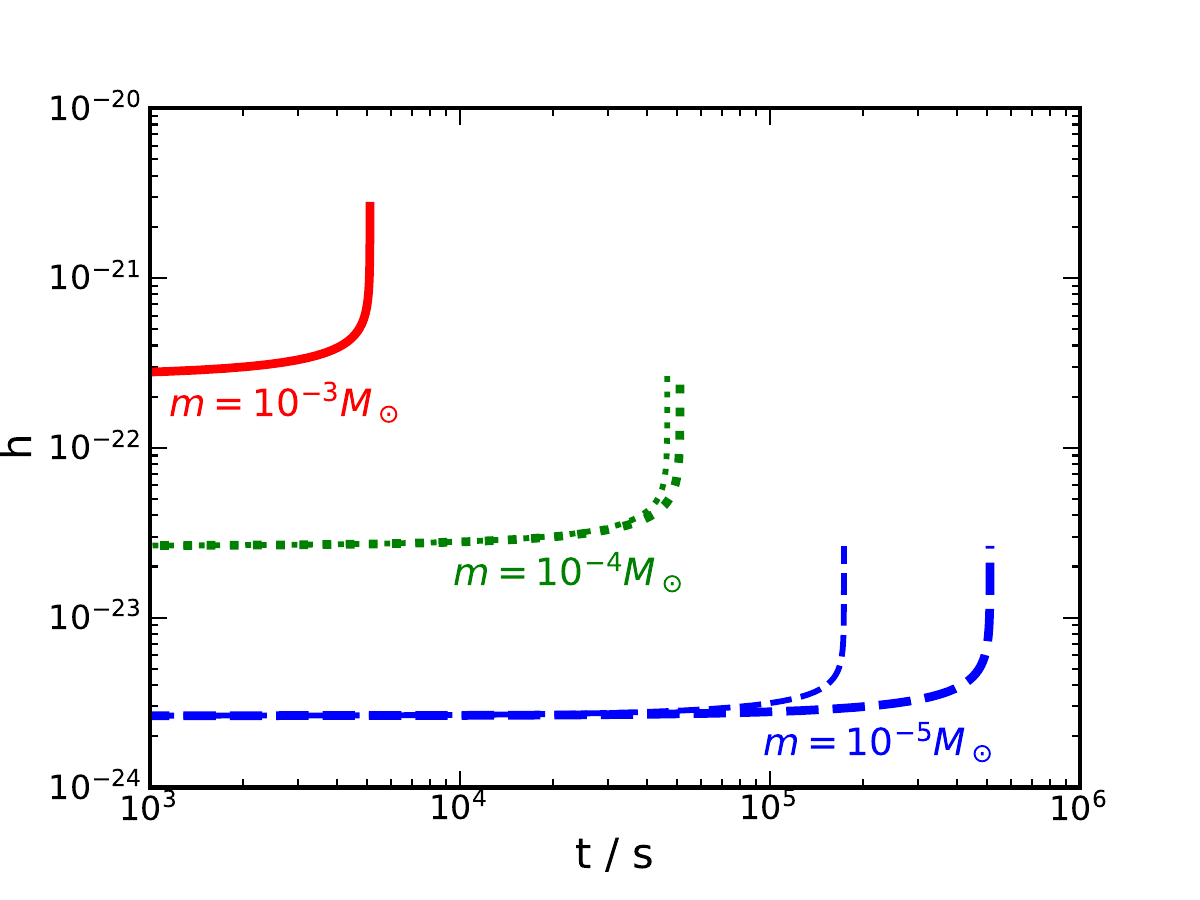}
    \end{minipage}
    \begin{minipage}{0.49\textwidth}
        \centering
        \includegraphics[width=\columnwidth]{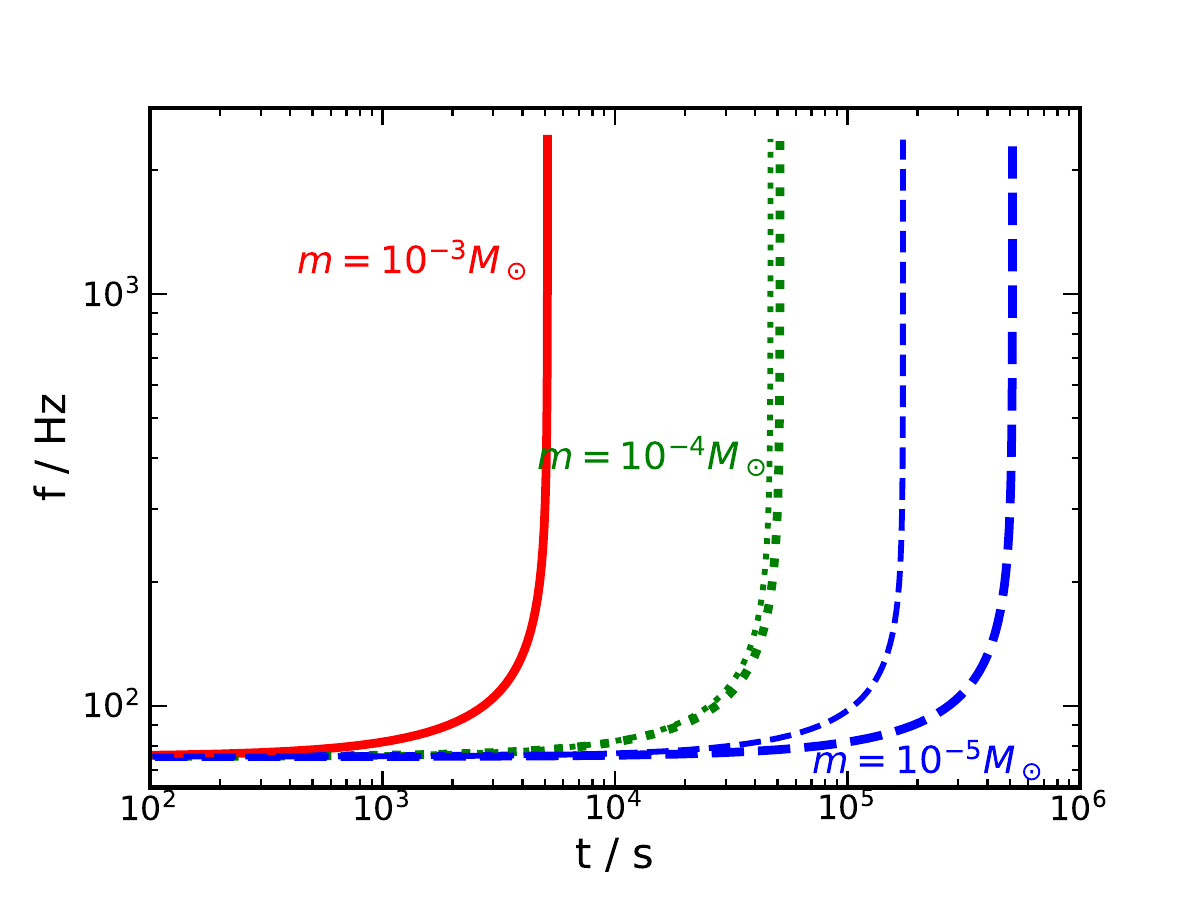}
    \end{minipage}
    \caption{Evolution of the GWs emitted from merging strange
        star-strange planet systems which are assumed to be face-on. The upper panel shows the evolution of the effective strain amplitude ($h$) and the lower panel shows the evolution of the frequency ($f$). In each panel, the solid, dotted and dashed curves correspond to the strange planet mass of $m = 10^{-3}$, $10^{-4}$, and $10^{-5}\,M_\odot$, respectively. For the thick curves, the magnetic moment of the strange star is taken as zero ($\mu = 0$), while for the thin curves, the magnetic moment is assumed to be $\mu = 10^{33}\,\mathrm{G\,cm^3}$ (corresponding to a surface magnetic field of $\sim 10^{15}$ G for the strange star). Note that in the case of $m = 10^{-3} \,M_\odot$, the effect of the magnetic field is almost negligible so that the thick curve and
        the thin curve are essentially overlapped.}
    \label{fig1}
\end{figure}

\section{Results}
\label{sec:result}

Using the equations described above, we have solved the merging process of strange star-strange planet systems and computed GW emissions numerically. In our calculations, the mass of the strange star is fixed as $M = 1.4\,M_\odot$, while three masses are taken for the strange planet, i.e. $m = 10^{-3}$, $10^{-4}$, and $10^{-5}\,M_\odot$. The distance of the planetary system is taken as 10 kpc. The initial separation between the strange star and the planet is assumed to be $1.5 \times 10^7$ cm. To assess the effects of the magnetic field, we take the magnetic moment as $\mu = 10^{33}\,\mathrm{G\,cm^3}$, which corresponds to a strongly magnetized magnetar with a surface field of $ B \sim 10^{15}$ G \citep{Duncan&Thompson_1992_ApJ,Woods&Thompson_2006_arXivbook,Kaspi_2010,Mereghetti_2015_SSRv,Turolla_2015_RPPh,Kaspi&Beloborodov_2017_ARA&A}. The results are compared with those of the $\mu = 0$ ($B = 0$) cases.

The effective strain amplitude is $h=\sqrt{h_+^2+h_\times^2}$, which can be calculated by using Equation~(\ref{eq:GW}). The upper panel of Figure~\ref{fig1} illustrates the evolution of $h$ during the inspiral process for face-on systems ($\iota=0$). We see that for a more massive strange planet, the system will merge more quickly and the GW emission is correspondingly stronger. The magnetic interaction accelerates the merging process, especially for the light strange planet cases. The evolution of the GW frequency ($f=2/T$) is plotted in the lower panel of Figure~\ref{fig1}. The frequency is mainly $\sim 100$ -- 1000 Hz, which falls in the frequency range of most ground-based GW detectors. Again, we see that the existence of the magnetic field speeds up the merging process.

\begin{figure*}
    \includegraphics[width=.8\textwidth]{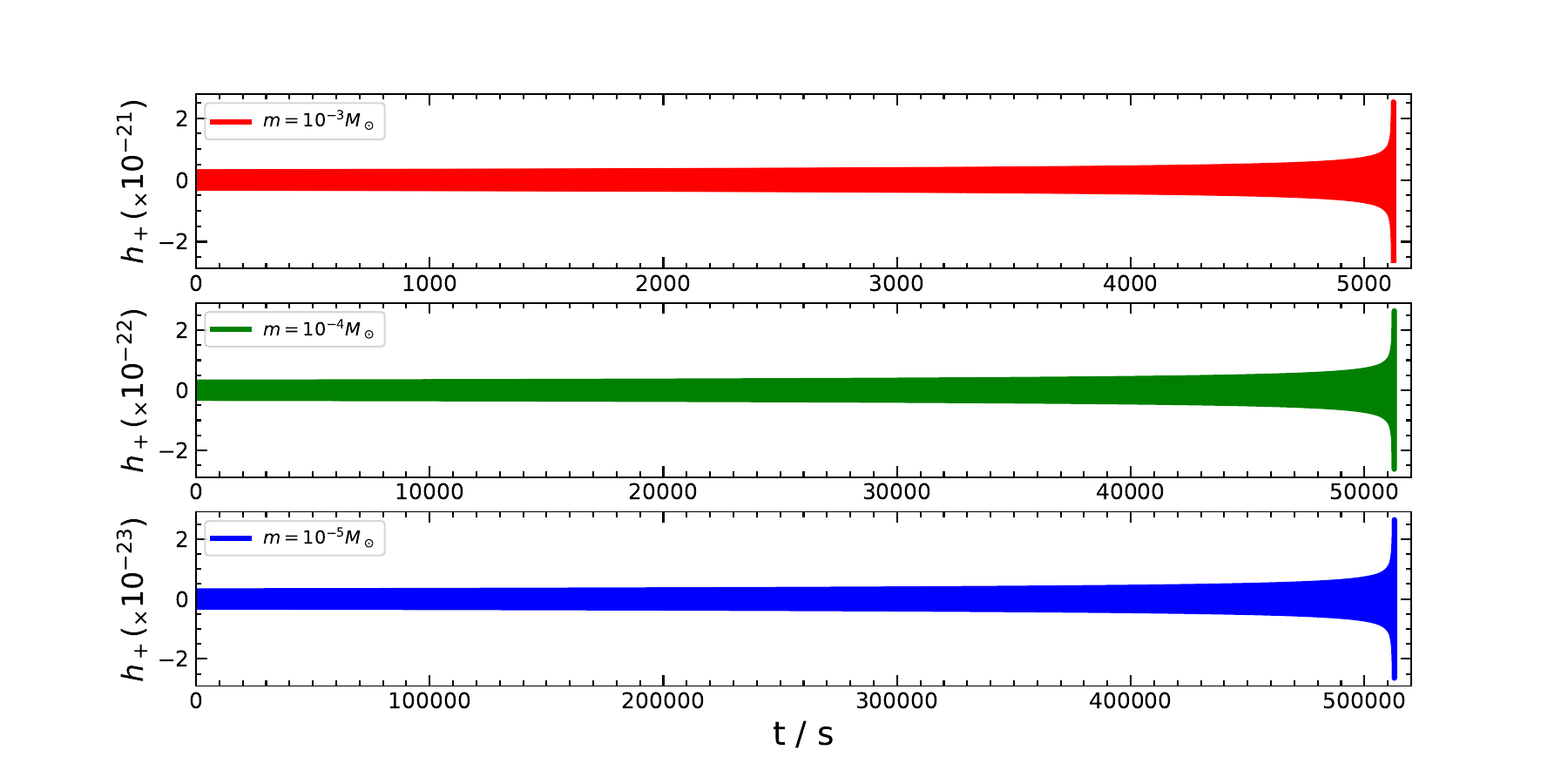}
    \centering
    \caption{Evolution of the GW profile (the $h_+$ component)
        for merging strange star-strange planet systems. The systems are assumed to be face-on and no magnetic interactions are considered ($\mu=0$). Three
        different masses are assumed for the strange planet,
        i.e. $ m = 10^{-3}, 10^{-4}$, and $10^{-5}\,M_\odot$, respectively.}
    \label{fig2}
\end{figure*}

\begin{figure*}
    \includegraphics[width=\textwidth]{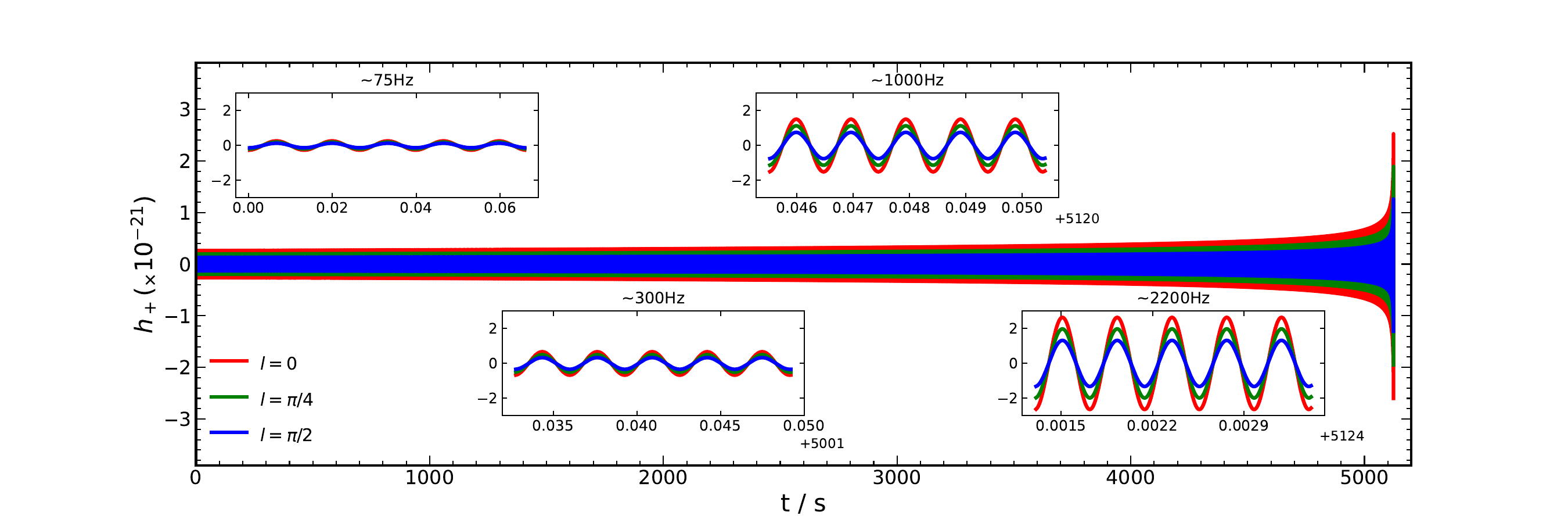}
    \centering
    \caption{Effects of the inclination angle on the GW profile
        (the $h_+$ component). Three different inclination angles, i.e. $\iota=0$ (face-on), $\iota=\uppi/4$, and $\iota=\uppi/2$ (edge-on), are shown in the plot. Here we take the magnetic moment of the strange star as $\mu=0$. The mass of the strange planet is $m = 10^{-3}\,M_\odot$. The insets show a zoom in of the GW profile at four moments, which correspond to a frequency of $\sim 75$, 300, 1000, and 2200 Hz, respectively. }
    \label{fig3}
\end{figure*}

Figure~\ref{fig2} shows the evolution of the ``plus'' polarization
GW waveforms for merging strange star-strange planet systems. The
effect of magnetic interactions is not considered here ($\mu=0$),
and the systems are assumed to be face-on ($\iota=0$). In fact,
the ``envelopes'' of these waveforms are just those strain
amplitude curves shown in Figure~\ref{fig1}.
Note that both the X and Y axes are in linear scale
rather than in logarithmic scale, which can help illustrate the
evolution of the GW waveforms more directly and also facilitates a
clear comparison with the following cases including magnetic
interactions. From Figure~\ref{fig2}, we see that the coalescence
timescale is inversely correlated with the planet mass ($m$). For
example, when $m = 10^{-3}\,M_\odot$, the coalescence timescale is
$\sim 5.12 \times 10^3$ s (the top panel). It increases to $\sim
5.13 \times 10^4$ s for $m = 10^{-4}\,M_\odot$ (the middle panel),
and increases to $\sim 5.13 \times 10^5 $ s for $m =
10^{-5}\,M_\odot$ (the bottom panel). At the same time, the strain
amplitude is proportional to $m$. These behaviors are consistent
with Equation~(\ref{eq:GW}). The effects of inclination angle on
the GW waveforms are illustrated in Figure~\ref{fig3}. Three
different inclinations, i.e. $\iota=0$ (face-on), $\iota=\uppi/4$,
and $\iota=\uppi/2$ (edge-on), are considered here as typical
examples. The effect of magnetic interactions is also not
included here ($\mu=0$). The insets show a zoom in of the GW
waveform at four typical moments, with the time marked on
the horizontal axis correspondingly. We see that the GW
amplitude becomes smaller as $\iota$ increases, and a face-on
observer would see the strongest GW emissions.

\begin{figure*}
    \includegraphics[width=.8\textwidth]{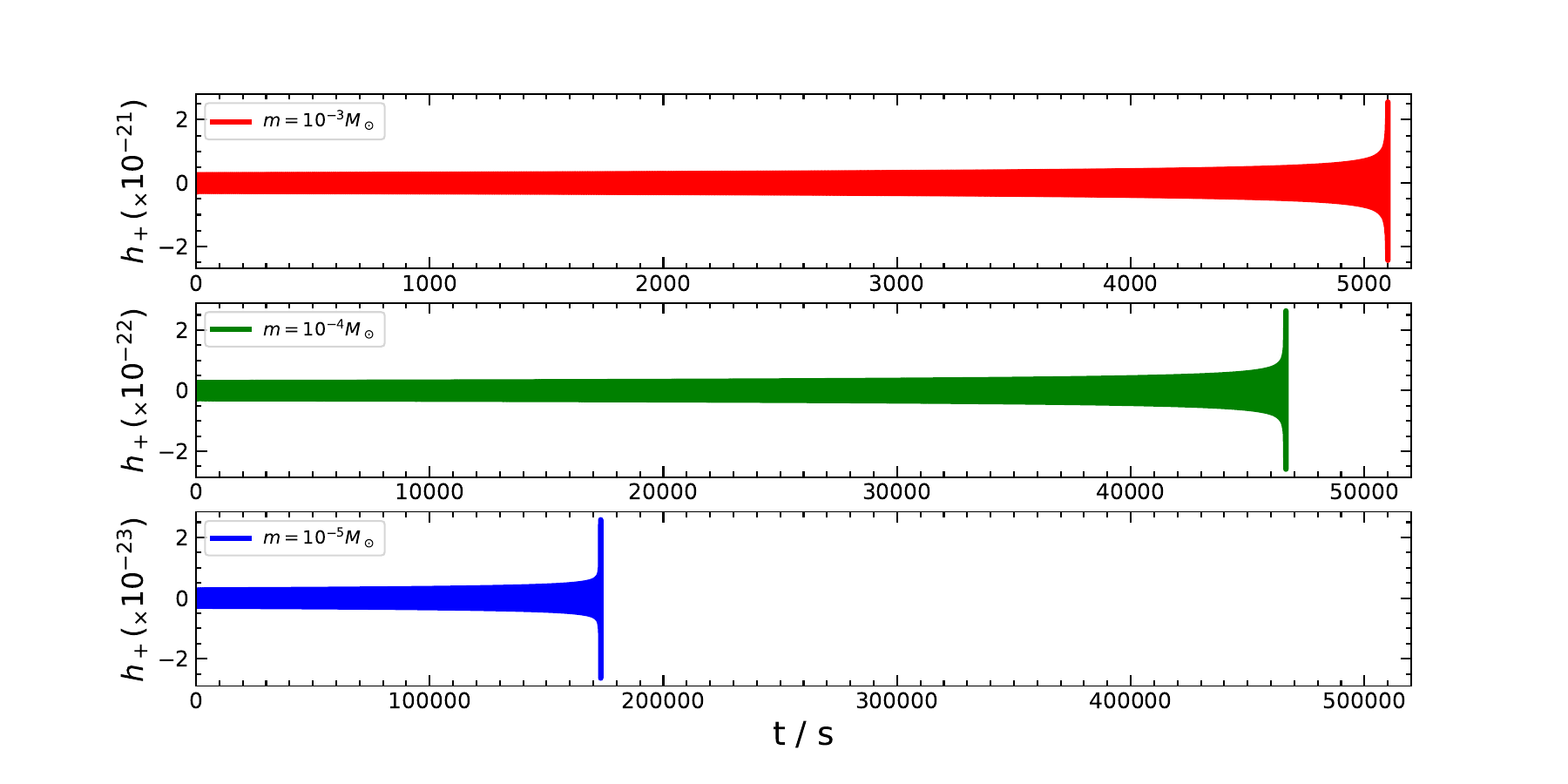}
    \centering
    \caption{Evolution of the GW profile (the $h_+$ component) for merging
        strange star-strange planet systems. The systems are assumed to be
        face-on. The magnetic moment of the strange star is taken
        as $\mu=10^{33}\,\mathrm{G\,cm^3}$. Three
        different masses are assumed for the strange planet,
        i.e. $ m = 10^{-3}, 10^{-4}$, and $10^{-5}\,M_\odot$, respectively.  }
    \label{fig4}
\end{figure*}

\begin{figure*}
    \includegraphics[width=\textwidth]{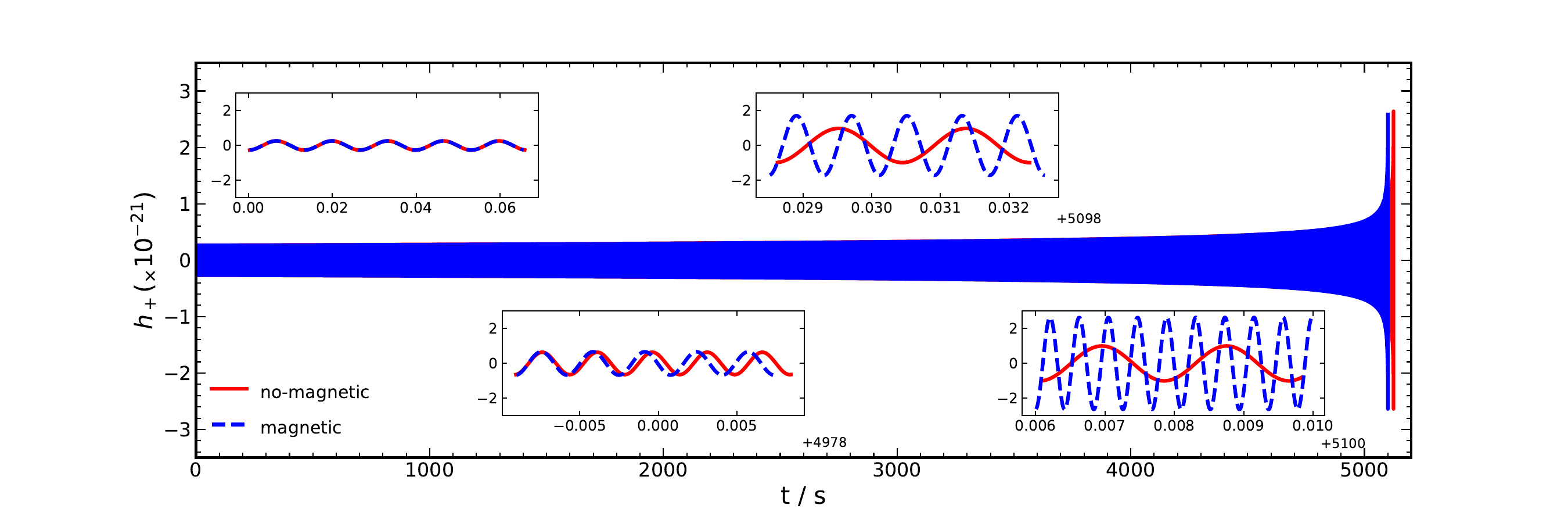}
    \centering
    \caption{A direct comparison of the GW waveforms for the cases with/without
        magnetic interactions. The systems are assumed to be face-on, and the
        mass of the strange planet is $m = 10^{-3}\,M_\odot$.
        The solid curve corresponds to the $\mu=0$ case and the dashed curve
        corresponds to a magnetic moment of $\mu=10^{33}\,\mathrm{G\,cm^3}$
        for the strange star. The insets show a zoom in of the GW profile
        at four exemplar moments.
    }
    \label{fig5}
\end{figure*}

Figure~\ref{fig4} shows the evolution of the $h_+$ component in
cases of strong magnetic interactions. We take a magnetic moment
of $\mu = 10^{33}\,\mathrm{G\,cm^3}$ for the strange star.
Comparing Figure~\ref{fig4} with Figure~\ref{fig2}, we
find that the magnetic field interaction significantly shortens
the coalescence timescale for the light strange planet cases. For
example, when $m = 10^{-5}\,M_\odot$, the coalescence timescale is
$\sim 5.13 \times 10^5 $ s in the $\mu = 0$ case (see the bottom
panel of Figure~\ref{fig2}), but it is $\sim 1.73 \times 10^5 $ s in
the $\mu = 10^{33}\,\mathrm{G\,cm^3}$ case (see the bottom panel
of Figure~\ref{fig4}). The timescale is reduced by a factor of
about two-thirds due to the magnetic interaction. On the other hand,
when the mass of the strange planet is larger, the effect becomes
less significant. For the planet mass of $m = 10^{-3}\,M_\odot$,
the coalescence timescale is $\sim 5.12 \times 10^3$ s in the $\mu
= 0$ case (see the top panel of Figure~\ref{fig2}), and it is
$\sim 5.10 \times 10^3$ s in the $\mu = 10^{33}\,\mathrm{G\,cm^3}$
case (see the top panel of Figure~\ref{fig4}). These two
timescales differ only slightly. The effect of magnetic field
can be easily understood by combining Equations~(\ref{eq:Emag})
and~(\ref{eq:Egw}): the energy dissipation rate due to GW emission
scales with the planet mass as $\dot{E}_\mathrm{GW}\propto m^2$,
while the magnetic interaction scales as
$\dot{E}_\mathrm{mag}\propto r^2\propto m^{2/3}$. Therefore, the
more massive the strange planet is, the less significant the
magnetic  interaction will be as compared with the GW dissipation.
Figure~\ref{fig5} presents a direct comparison of the GW waveforms
for the cases with/without magnetic interactions. One can see that
at early stages, the waveforms are almost identical for the two
cases. As time goes on, the frequency becomes slight different so
that the phase shift becomes obvious. At the final chirp stage,
the waveform is essentially quite different in the two cases. It
hints that the GW observations may potentially provide some useful
clues for measuring the magnetic field of such compact stars.

To successfully observe a GW signal, its strain spectral amplitude should be higher than the sensitivity curve of GW detectors. Ground-based detectors mainly operates in the frequency domain. We thus need to analyze the GW signals via the Fourier transform method. Adopting the stationary phase approximation, the Fourier transform of the $h_+$ component of GW emissions can be expressed as  \citep{Moore_2015_CQGra}
\begin{equation}
    \tilde{h}_+(f) \approx \frac{h_0}{\sqrt{2}}\int_{-\infty}^{+\infty}
    \left[ \exp \left( 2 \uppi\mathrm{i} \dot{f}t^2\right)
    + \exp \left(-2\uppi\mathrm{i} \dot{f}t^2 \right) \right]\mathrm{d}t
    \approx \frac{h_0}{\sqrt{2\dot{f}}},
\end{equation}
where
\begin{equation}
    h_0=\frac{h_+}{\sqrt{2}\cos2\varphi}.
\end{equation}
The $\tilde{h}_\times$ polarization component can also be calculated similarly. The overall average of the Fourier transform function is then
\begin{equation}
    \tilde{h}(f)=\sqrt{\left<\tilde{h}_+^2(f)\right>+\left<\tilde{h}_\times^2(f)\right>}.
\end{equation}
Using this expression, the GW strain spectral amplitude \citep[square root of the GW power spectral density;][]{Moore_2015_CQGra,2020RAA....20..137Z,Zou&Huang_2022} can finally be
derived as
\begin{equation}
    h_f=2f^{1/2}\left|\tilde{h}(f)\right|.
\end{equation}

\begin{figure}
    \includegraphics[width=\columnwidth]{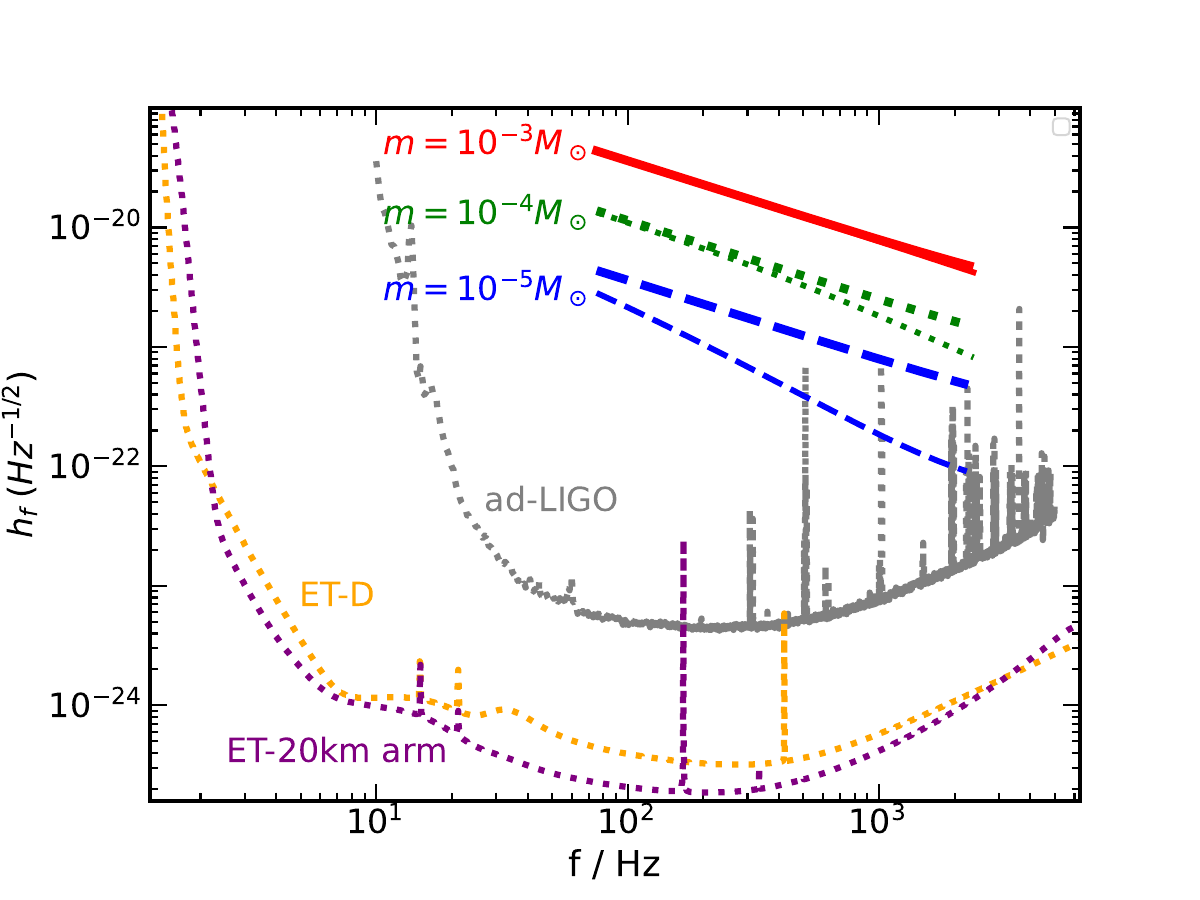}
    \caption{Strain spectral amplitude of GWs from various
        strange star-strange planet systems, plotted versus frequency.
        The planetary systems are assumed to be face-on, and the distance
        is 10 kpc. The mass of the strange planet is taken
        as $m = 10^{-3}$, $10^{-4}$ and $10^{-5}\,M_\odot$, respectively.
        For the thick curves, the magnetic moment of the strange star is
        taken as zero ($\mu = 0$), while for the thin curves, the magnetic
        moment is assumed to be $\mu = 10^{33}\,\mathrm{G\,cm^3}$. The
        sensitivity curves of the ad-LIGO O3, ET-D (Einstein Telescope
        designed in 2011), and ET-20km arm (the updated Einstein Telescope
        with 20 km arms designed in 2021) are also plotted for a direct comparison. }
    \label{fig6}
\end{figure}

Figure~\ref{fig6} plots the strain spectral amplitude of various face-on strange star-strange planet systems. Generally, the amplitude is higher when the planet mass is larger. In the $\mu = 0$ cases, the $\lg h_f$ -- $\lg f$ plots are essentially straight lines with a universal slope of $-2/3$, which is consistent with the analytical expectations \citep{Moore_2015_CQGra}. When the magnetic interactions take effect, the $\lg h_f$ -- $\lg f$ plots deviate from the straight lines obviously. The deviation is more significant when the planet mass is smaller. It further clearly shows that the magnetic field of the compact star leads to different behavior of the GW emission, which could potentially be used to measure the magnetic field strength of compact stars.

The sensitivity curves of the advanced Laser Interferometer
Gravitational-wave
Observatory\footnote{\url{https://dcc-lho.ligo.org/LIGO-T2000012/public}}
\citep[ad-LIGO, O3 stage;][]{Harry_2010}, the future Einstein
Telescope designed in 2011
\footnote{\url{https://www.et-gw.eu/index.php/etsensitivities}}
\citep[ET-D;][]{Hild_2008arXiv,2011CQGra..28i4013H,Maggiore_2020_JCAP}
and the updated future Einstein Telescope designed in 2021
with the length of arms taken as 20 km
\footnote{\url{https://apps.et-gw.eu/tds/ql/?c=16492}}\citep[ET-20km
arm;][]{Branchesi_2023_JCAP} are also plotted in
Figure~\ref{fig6} for a direct comparison. We see that the
amplitudes are generally much higher than the sensitivity curves,
thus such a kind of GW signals could potentially be detected by
current and future GW experiments. In fact, GW emissions from
merging strange star-strange planets in local galaxies up to a
distance of several Mpc could also be detectable
\citep{Geng_2015_Apjl}. Signals from such events with an extreme
mass ratio ($> 10^3$ -- $10^4$) are suggested to be paid special
attention in current and future GW observations.

\section{Conclusions and Discussion}
\label{sec:ending}

In this study, the inspiral and merging process of a strange planet with respect to its host strange star is investigated, aiming to provide detailed GW waveform information and GW templates for detecting such events by using current and future GW experiments. The effect of the inclination angle is considered and the magnetic interaction is included. It is found that a strong magnetic field of the strange star can markedly speed up the inspiral process, leading to a much shorter coalescence timescale. The effect of magnetic interactions is more significant for the less massive strange planet cases. By comparing the strain spectral amplitude of GWs with the sensitivity curves of ad-LIGO and ET, it is shown that the merging events occurring in our Galaxy can be detected by these GW experiments. The effects of magnetic interactions can be discerned from GW observations, hopefully providing a new method to measure the magnetic field of compact stars.

It is worth mentioning that primordial black holes (PBH) can be captured by compact stars and form close-in planetary systems \citep{Genolini_2020}. In these systems, when the planetary-mass PBH finally merges with the compact star, strong GW emissions will also be generated, whose characteristics should be very similar to the cases studied here. Still we can hopefully distinguish the strange star-strange planet mergers from the compact star-PBH mergers by considering their different behaviors in the ring down stage. After a PBH comes into the surface of its host, it will keep tunneling inside the compact star and inspiral towards the center. In this process, a strong ring down signal will be generated in the GW emission, which can be detected by ad-LIGO at a high confidence level \citep{Zou&Huang_2022}. Additionally, a compact star-PBH merger will end up with a strong electromagnetic outburst, since the compact star will finally be swallowed by the PBH. By contrast, a strange planet will finally collide and coalesce with the host strange star, producing relative shorter and much weaker ring down signals.

Close-in planets are usually found to be tidally locked \citep{2018exha.book.....P}. However, in our study, such a tidal effect has not been considered. This is acceptable because the tidal deformability of strange planets are generally very small \citep{2021PhRvD.104l3028W}. Anyway, at the final stage of the merging process, tidal deformability may still take effect on the GW waveform when the strange planet approaches the surface of the strange star. This effect may need to be further addressed in future studies.

Targeted filter-matched searches in archival GW data have already yielded many interesting results \citep[e.g.,][]{2018CQGra..35c5016N,2021ApJ...912...53W,2024MNRAS.tmp...92W}. With the GW templates available for merging strange planetary systems, an in-depth search for GW signals corresponding to these extreme mass ratio merging events in currently available GW data as well as in future GW experiments is solicited, which will help test the SQM hypothesis and probe the nature of dense nuclear matter.

\section*{Acknowledgments}

Z.-C Z. gratefully acknowledges An-Dong Chen for enlightening
discussion on exoplanet detection. This study was supported
by the National Natural Science Foundation of China (Grant Nos. 12233002, 11988101, U2031117),
by the National SKA Program of China Nos. 2020SKA0120300, 2020SKA0120200 and No. 2022SKA0120102,
by the National Key R\&D Program of China (2021YFA0718500,  2023YFE0102300),
by the CAS ``Light of West China'' Program (No. 2021-XBQNXZ-005), and by Xinjiang Tianshan Talent Program.
YFH also acknowledges the support from the Xinjiang Tianchi Program.


\section*{Data Availability}

No new data were generated or analysed in support of this research.



\bibliographystyle{mnras}
\bibliography{ref} 






\bsp    
\label{lastpage}
\end{document}